\documentclass[aps,prb,reprint,showpacs,superscriptaddress,longbibliography]{revtex4-1}

\usepackage{graphicx}
\usepackage{physics}

\usepackage{amsmath,amssymb}
\usepackage{color}
\usepackage[normalem]{ulem}

\usepackage{siunitx}
\usepackage{amsmath,amssymb}
\usepackage{color}
\usepackage{romannum}

\usepackage{bm}
\usepackage{hyperref}
\hypersetup{
setpagesize=false,
colorlinks=true,
linkcolor=blue,
citecolor=red,
}

\begin{document}
\title{Valley Hall Viscosity in Gapped Graphene with and without a Magnetic Field}

\author{Danyu Shu}
\email{shudanyu24@mails.ucas.ac.cn}
\affiliation{%
Kavli Institute for Theoretical Sciences, University of Chinese Academy of Sciences, Beijing, 100190, China.
}%

\author{Hiroshi Funaki}
\affiliation{%
Center for Spintronics Research Network, Keio University, Yokohama 223-8522, Japan
}%
\affiliation{%
Kavli Institute for Theoretical Sciences, University of Chinese Academy of Sciences, Beijing, 100190, China.
}%
\author{Ai Yamakage}
\affiliation{
Department of Physics, Nagoya University, Nagoya 464-8602, Japan
	}
\author{Ryotaro Sano}
\affiliation{%
Department of Physics, Kyoto University, Kyoto 606-8502, Japan
}%
\affiliation{Institute of Solid State Physics, University of Tokyo, Kashiwa 277-8581, Japan}

\author{Mamoru Matsuo }
\email{mamoru@ucas.ac.cn}
\affiliation{%
Kavli Institute for Theoretical Sciences, University of Chinese Academy of Sciences, Beijing, 100190, China.
}%

\affiliation{%
CAS Center for Excellence in Topological Quantum Computation, University of Chinese Academy of Sciences, Beijing 100190, China
}%
\affiliation{%
RIKEN Center for Emergent Matter Science (CEMS), Wako, Saitama 351-0198, Japan
}%
\affiliation{%
Advanced Science Research Center, Japan Atomic Energy Agency, Tokai, 319-1195, Japan
}%

\date{\today}

\begin{abstract}
Hall viscosity is conventionally defined through the stress response to time-dependent strain, a perturbation that is difficult to implement in solid-state experiments.
We formulate a related viscoelastic response to static, spatially inhomogeneous electric fields and compare it with the strain-based response.
For gapped graphene in a perpendicular magnetic field, the two formulations give the same Landau-level response, whose Hall viscosity is asymmetric between the two valleys.
At zero magnetic field, a valley-even quantum-metric coefficient combines with the valley-odd Hall conductivity to produce equal and opposite valley-resolved responses; global time-reversal symmetry therefore forces the net Hall viscosity to vanish.
In an insulating state, exact particle--hole symmetry eliminates this zero-field response, whereas particle--hole-symmetry breaking generates a cutoff-dependent geometric contribution from the occupied Fermi sea.
These results connect valley-dependent viscoelasticity with electromagnetic response in gapped Dirac materials and clarify the conditions under which a valley Hall viscosity can arise.

\end{abstract}

\maketitle

\section{Introduction} 
Hydrodynamic electron flow~\cite{gurzhiElectronElectronCollisionsNew1995,r.n.gurzhiHydrodynamicEffectsSolids1968} arises when momentum-conserving electron--electron scattering dominates over momentum-relaxing processes in ultraclean two-dimensional (2D) systems. Experiments in high-mobility 2D electron gases~\cite{m.j.m.dejongHydrodynamicElectronFlow1995,aydincemkeserGeometricControlUniversal2021,j.goothThermalElectricalSignatures2018}, graphene~\cite{d.a.bandurinNegativeLocalResistance2016,andrewlucasHydrodynamicsElectronsGraphene2018,philipj.w.mollEvidenceHydrodynamicElectron2016}, and related van der Waals materials have identified viscous electron transport through nonlocal measurements and direct imaging~\cite{josepha.sulpizioVisualizingPoiseuilleFlow2019,markj.h.kuImagingViscousFlow2020,a.aharon-steinbergDirectObservationVortices2022}. These observations demonstrate that viscosity can govern electronic flow in mesoscale devices.

When an external magnetic field breaks time-reversal symmetry ($\mathcal{T}$), an electron fluid can exhibit dissipationless odd, or Hall, viscosity~\cite{j.e.avronViscosityQuantumHall1995,j.e.avronOddViscosity1998}. Hall viscosity has been studied as a geometric and, in certain settings, topological response~\cite{boyangEquivalenceChernBands2025,x.g.wenShiftSpinVector1992,haochenMagicNonlocalGeometric2025,carolinapaivaGeometricalResponsesGeneralized2025,ioannismatthaiakakisTorsionalHallViscosity2025} closely related to the geometry of quantum Hall states~\cite{barrybradlynKuboFormulasViscosity2012,michaelp.zaletelTopologicalCharacterizationFractional2013,bendeguzoffertalerViscoelasticResponseQuantum2019,n.readHallViscosityOrbital2011,n.readNonAbelianAdiabaticStatistics2009,takuyafurusawaHallViscosityAphase2021}. Odd-viscous signatures have also been inferred experimentally at moderate magnetic fields~\cite{a.i.berdyuginMeasuringHallViscosity2019,lucav.delacretazTransportSignaturesHall2017}. The conventional definition, however, involves the response of the stress tensor to time-dependent strain, which is difficult to control and measure in solid-state devices.

Here we formulate a viscoelastic response to static, spatially inhomogeneous electric fields, motivated by established relations between viscosity and spatially dispersive conductivity~\cite{carloshoyosHallViscosityElectromagnetic2012}. We apply this construction to gapped graphene, a 2D Dirac material with two valleys, $K$ and $K'$, and benchmark it against the conventional strain-based Kubo response in the Landau-level regime. The calculation identifies two manifestations of \textit{valley Hall viscosity}. In a perpendicular magnetic field, global $\mathcal{T}$ breaking produces a valley asymmetry in the Landau-level Hall viscosity~\cite{mohammadsherafatiHallViscosityNonlocal2019,selchValleyHallViscosity2026}. At zero magnetic field, a valley-even quantum-metric response combines with the valley-odd Hall conductivity. The resulting valley-resolved viscosities are equal and opposite, as in the valley Hall effect~\cite{dixiaoValleyContrastingPhysicsGraphene2007,y.shimazakiGenerationDetectionPure2015,jieunleeElectricalControlValley2016,k.f.makValleyHallEffect2014}, whereas the net macroscopic response vanishes by $\mathcal{T}$ symmetry. In an insulating state, exact particle--hole symmetry eliminates this zero-field response; breaking that symmetry generates a cutoff-dependent geometric contribution from the occupied Fermi sea.

The remainder of the paper is organized as follows. Section~\ref{Formalism} introduces the viscosity tensor and compares the strain-based calculation (Sec.~\ref{Strain Deformation}) with the inhomogeneous-electric-field construction (Sec.~\ref{Inhomogeneous Electric Field}). Section~\ref{Valley Hall Viscosity with magnetic field} evaluates the Landau-level Hall viscosity of gapped graphene using both methods and identifies its valley asymmetry. Section~\ref{Valley Hall Viscosity without magnetic field} separates the Fermi-surface and Fermi-sea contributions to the zero-field valley response.

\section{Formalism}
\label{Formalism}
\begin{figure}[htbp]
  \centering
  \includegraphics[width=8cm]{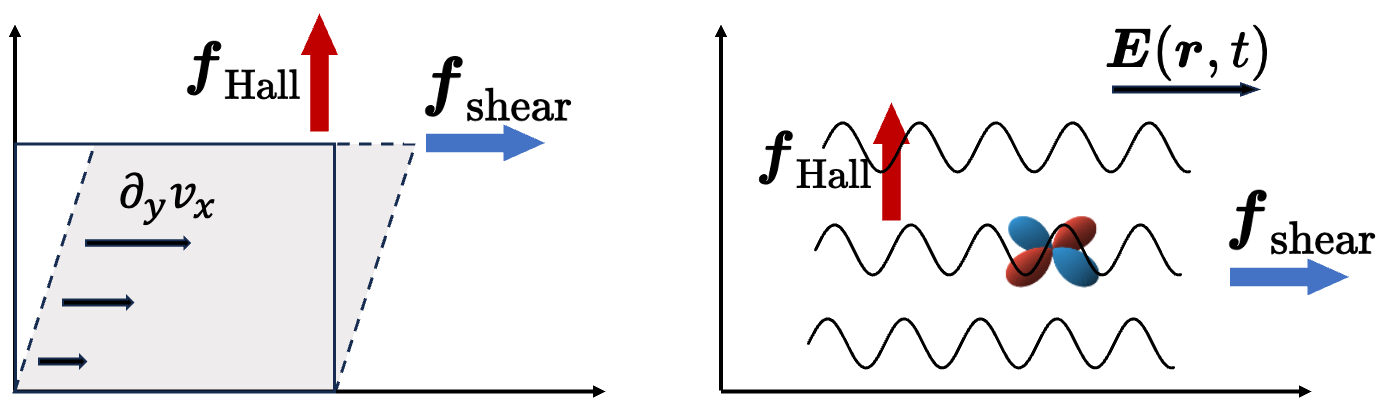}
  \caption{Schematic of viscous responses in an electron fluid. (a) Viscous forces induced by a strain-rate gradient. The shear force ($\boldsymbol{f}_{\text{shear}}$, blue arrow) is parallel to the deformation rate, whereas the Hall-viscous force ($\boldsymbol{f}_{\text{Hall}}$, red arrow) is perpendicular to it. (b) The gradient of a spatially inhomogeneous electric field couples to the electronic quadrupole moment and generates a macroscopic viscous force.}
  \vspace{-1em}
  \label{fig:schematic}
\end{figure}

In the hydrodynamic regime, momentum transport is characterized by the viscosity tensor $\eta_{ij;kl}$. Linear-response theory defines this tensor through the response of the stress expectation value $\langle\hat{T}_{ij}\rangle$ to the strain-rate tensor $\dot{\lambda}_{kl} \equiv \partial_t \lambda_{kl}$~\cite{j.e.avronOddViscosity1998,j.e.avronViscosityQuantumHall1995}:
\begin{align}
    \label{def:viscosity}
    \langle\hat{T}_{ij}\rangle=-\eta_{ij;kl}\frac{\partial \lambda_{kl}}{\partial t}.
\end{align}
For an isotropic two-dimensional (2D) system, spatial symmetry reduces the viscosity tensor to three independent scalar coefficients:
\begin{multline}
\label{three components}
\eta_{ij;kl}=\zeta\delta_{ij}\delta_{kl}+\eta_S(\delta_{ik}\delta_{jl}+\delta_{il}\delta_{jk}-\delta_{ij}\delta_{kl})\\-\eta_H(\delta_{jk}\epsilon_{il}-\delta_{il}\epsilon_{kj}).
\end{multline}
Here, $\zeta$ and $\eta_S$ are the bulk and shear viscosities, respectively. The antisymmetric coefficient $\eta_H$ is the Hall viscosity. A nonzero macroscopic charge Hall viscosity requires broken $\mathcal{T}$. With the force-density convention $f_j=-\partial_iT_{ij}$, this decomposition gives the following contributions~\cite{etienneguyonPhysicalHydrodynamics2015,lucianorezzollaRelativisticHydrodynamics2018} (Fig.~\ref{fig:schematic}):
\begin{equation}
    \begin{aligned}
        \boldsymbol{f}_{\mathrm{bulk}}&=\zeta\nabla(\nabla\cdot\boldsymbol{v}),\\
        \boldsymbol{f}_{\mathrm{shear}}&=\eta_S\nabla^2\boldsymbol{v},\\
        \boldsymbol{f}_{\mathrm{Hall}}&=\eta_H\nabla^2\boldsymbol{v}\times\hat{\boldsymbol{z}}.
    \end{aligned}
\end{equation}
The velocity field $\boldsymbol{v}(\boldsymbol{x},t)$ is related to the strain tensor through
\begin{align}
    \label{v-lambda}
    \frac{\partial\lambda_{ij}}{\partial t}=\frac{1}{2}\left(\frac{\partial v_i}{\partial x_j}+\frac{\partial v_j}{\partial x_i}\right).
\end{align}
Nonnegative entropy production requires the longitudinal coefficients $\zeta$ and $\eta_S$ to be nonnegative, whereas the dissipationless transverse coefficient $\eta_H$ is not constrained in sign. We use a fixed orientation convention throughout; reversing the carrier type, valley chirality, or magnetic-field direction can reverse the sign of $\eta_H$.

We first review the conventional derivation of the viscosity tensor from an explicit strain deformation and a stress--stress correlation function. We then derive the viscoelastic response to a spatially inhomogeneous electric field from a stress--velocity correlation function, without an explicit strain coupling.

\subsection{Strain Deformation}
\label{Strain Deformation}
Microscopically, the stress-tensor operator $\hat{T}_{ij}$ is the time derivative of the strain generator $\hat{J}_{ij}$. The generator obeys the canonical commutation relations with the position and momentum operators~\cite{barrybradlynKuboFormulasViscosity2012,f.d.m.haldaneIncompressibleQuantumHall2023}:
\begin{align}
    \label{generator}
    [\hat{J}_{ij},\hat{x}_k]=i\delta_{jk}\hat{x}_i, \qquad [\hat{J}_{ij},\hat{p}_k]=-i\delta_{ik}\hat{p}_j.
\end{align}

A macroscopic strain $\lambda_{\mu\nu}$ enters the perturbation Hamiltonian in either of two gauge-equivalent forms. The first couples the strain directly to the stress tensor,
\begin{align}
\label{gauge1}
    H^\prime=-\lambda_{\mu\nu}\hat{T}_{\mu\nu},
\end{align}
whereas the second couples the strain rate to the strain generator:
\begin{align}
    \label{gauge2}
    H^\prime=-\frac{\partial\lambda_{\mu\nu}}{\partial t}\hat{J}_{\mu\nu}.
\end{align}
We use the first gauge choice, Eq.~\eqref{gauge1}. The corresponding stress response includes the contact term
\begin{align}
    \label{contact}
    \hat{T}^c_{ij}=-i\lambda_{kl}[\hat{J}_{kl},\hat{T}_{ij}],
\end{align}
which is analogous to the diamagnetic-current term in an electromagnetic response.

The Kubo formula~\cite{r.kuboStatisticalMechanicalTheoryIrreversible1957,d.a.greenwoodBoltzmannEquationTheory1958} expresses the frequency-dependent viscosity tensor as the sum of the stress--stress correlation function and the contact contribution:
\begin{align}
    \label{T-T}
    \eta_{ij;kl}(\omega)=\frac{1}{\omega^+}\Bigg(&\langle[\hat{T}_{ij},\hat{J}_{kl}]\rangle_0 \notag\\
    &+\int_0^\infty\mathrm{d}t\,e^{i\omega^+t}\langle[\hat{T}_{ij}(t),\hat{T}_{kl}]\rangle_0\Bigg).
\end{align}
Here, $\omega^+ \equiv \omega + i0^+$, and the unperturbed Hamiltonian $\hat{H}_0$ determines the operator time dependence, $\hat{A}(t)=\exp(i\hat{H}_0t)\hat{A}\exp(-i\hat{H}_0t)$. For a noninteracting system, expansion in single-particle eigenstates $|n\rangle$ with energies $\epsilon_n$ yields
\begin{align}
    \label{T-T_corre}
    \eta_{ij;kl}(\omega)=-i\sum_{m\neq n}\bra{m}\hat{T}_{ij}\ket{n}\bra{n}\hat{T}_{kl}\ket{m}F_{mn}(\omega)
\end{align}
where the occupation factors and energy denominators are collected in
\begin{align}
    F_{mn}(\omega)=\frac{f(\epsilon_m)-f(\epsilon_n)}{(\epsilon_m-\epsilon_n)(\epsilon_m-\epsilon_n+\omega^+)},
\end{align}
and $f(\epsilon)$ is the Fermi--Dirac distribution.

\subsection{Inhomogeneous Electric Field}
\label{Inhomogeneous Electric Field}
We now formulate the viscoelastic response to the gradient of a spatially inhomogeneous electric field without introducing an explicit lattice-strain coupling. In linear response, the stress tensor satisfies $T_{ij} = \chi_{ij;kl}\partial_k E_l$. For the symmetric, parity-even part of the response in an isotropic 2D system, spatial symmetry reduces $\chi_{ij;kl}$ to two scalar coefficients, $\chi$ and $t$:
\begin{align}
\label{chi}
    \chi_{ij;kl}=\chi(\delta_{ij}\delta_{kl}-\delta_{ik}\delta_{jl}-\delta_{il}\delta_{jk})-\frac{t}{2}\delta_{ij}\delta_{kl}.
\end{align}
The induced stress tensor is therefore
\begin{align}
    \label{Tflexo}
    T_{ij}=\left(\chi-\frac{t}{2}\right) \delta_{ij}\nabla\cdot\boldsymbol{E}-\chi\left(\frac{\partial E_i}{\partial x_j}+\frac{\partial E_j}{\partial x_i}\right).
\end{align}
The generalized Ohm's law relates this electrical response to the fluid velocity $\boldsymbol{v}$. Let $\rho_{ij}$ denote the resistivity tensor and write the current density as $j_k=-e\bar{n}v_k$, where $\bar{n}$ is the carrier density and $-e$ is the electron charge. The electric field is then $E_i=-e\bar{n}\rho_{ik}v_k$. Substituting this relation into $f_j=-\partial_iT_{ij}$ gives
\begin{align}
    \label{force}
    \boldsymbol{f}={}&\frac{t\kappa_1}{2}\nabla(\nabla\cdot\boldsymbol{v})
    +\chi\kappa_1\nabla^2\boldsymbol{v} \notag\\
    &+\chi\kappa_2\nabla^2\boldsymbol{v}_\perp
    +\frac{t\kappa_2}{2}\nabla(\nabla\cdot\boldsymbol{v}_\perp).
\end{align}
Here, $\boldsymbol{v}_\perp \equiv \boldsymbol{v}\times\hat{\boldsymbol{z}}$, $\kappa_1=-e\bar{n}\rho_{xx}$, and $\kappa_2=e\bar{n}\rho_{xy}$. Comparison with the Navier--Stokes viscous force identifies the bulk ($\zeta$), shear ($\eta_S$), and Hall ($\eta_H$) coefficients:
\begin{align}
\label{3coefficient}
    \zeta=\frac{t\kappa_1}{2}, \qquad \eta_S=\chi\kappa_1, \qquad \eta_H=\chi\kappa_2.
\end{align}

The final term in Eq.~\eqref{force} is proportional to the gradient of the fluid vorticity because $\nabla\cdot\boldsymbol{v}_\perp=\partial_xv_y-\partial_yv_x$. This term is absent from the conventional isotropic Navier--Stokes form. The coefficient $t$ controls the trace response and the additional vorticity-gradient term; neither term enters the Hall coefficient $\eta_H$ considered below. Because the dissipative shear viscosity depends strongly on disorder and momentum-relaxing scattering, we focus on $\eta_H$, which is determined by $\chi$ and the off-diagonal resistivity through $\kappa_2$.
The electric-field construction instead involves a stress--velocity correlator. A momentum derivative extracts the spatial gradient of the applied field, giving
\begin{align}
    \label{T-V_corre}
    \eta_H(\omega)={}&e\hbar\kappa_2\frac{\partial}{\partial q_x}
    \sum_{m\neq n}\bra{m}\hat{T}_{xy}(\boldsymbol{q})\ket{n}\notag\\
    &\times\bra{n}\hat{v}_{y}(-\boldsymbol{q})\ket{m}
    F_{mn}(\omega)\Big|_{q=0}.
\end{align}
where $\hat{T}_{ij}(\boldsymbol{q})$ and $\hat{v}_{y}(-\boldsymbol{q})$ are Fourier components of the stress and velocity operators, respectively.

This construction expresses the Hall coefficient as the product of the spatially dispersive stress coefficient $\chi$ and a transverse transport coefficient encoded in $\rho_{xy}$. The coefficient $\chi$ depends on the local band geometry and need not require macroscopic $\mathcal{T}$ breaking. The transverse response supplies the $\mathcal{T}$-odd factor needed for a net charge Hall viscosity. In a globally $\mathcal{T}$-invariant multivalley system, the Hall conductivities of opposite valleys cancel, and the net macroscopic response vanishes. A valley-resolved response may nevertheless remain finite when intervalley relaxation is sufficiently weak that the two valley channels can be treated separately. In the zero-field calculation below, $\chi$ is valley even, whereas the valley Hall conductivity is valley odd.

Related forms of Eq.~\eqref{T-V_corre} have been derived from Ward identities or momentum conservation~\cite{edwardtaylorViscosityStronglyInteracting2010,biaohuangHallViscosityRevealed2015}. Here we evaluate the stress response to a nonuniform electric field directly and compare it with the strain-based result in the Landau-level regime.

\section{Valley Hall Viscosity with a Magnetic Field}

\label{Valley Hall Viscosity with magnetic field}
We first apply the inhomogeneous-electric-field construction to monolayer gapped graphene in a perpendicular magnetic field. The field breaks global $\mathcal{T}$ and quantizes the spectrum into Landau levels (LLs), allowing an analytical comparison with the strain-based calculation. Inversion-symmetry breaking distinguishes the $K$ and $K'$ valleys of the massive Dirac model. In particular, valley--sublattice locking ties the energy and geometric character of the zeroth LL ($n=0$) to the valley index. The LL Hall viscosity is therefore valley dependent. Agreement between the two calculations provides an explicit benchmark for the electric-field construction in this model and regime.

\subsection{Model}
\label{Model}
The model Hamiltonian for monolayer gapped graphene~\cite{c.l.kaneQuantumSpinHall2005,yankowitzEmergenceSuperlatticeDirac2012,n.r.cooperThermoelectricResponseInteracting1997} in a perpendicular magnetic field $B$ is
\begin{equation}
    \label{model}
    H^\xi=v_F(\xi\pi_x\sigma_x+\pi_y\sigma_y)+\Delta\sigma_z,
\end{equation}
where $\boldsymbol{\pi}=-i\hbar\boldsymbol{\nabla}+e\boldsymbol{A}$, $\boldsymbol{\nabla}\times\boldsymbol{A}=B\hat{\boldsymbol{z}}$, $\xi=\pm1$ labels the two valleys, and $\boldsymbol{\sigma}$ denotes the Pauli matrices acting on the sublattice degree of freedom. We take $B>0$ throughout this section; the response for the opposite field direction follows by time reversal. The continuum Dirac approximation applies when the magnetic length $l_B=\sqrt{\hbar/(eB)}$ is much larger than the lattice constant $a$.

The eigenvalues of Eq.~\eqref{model} are
\begin{subequations}
    \begin{align}
        \epsilon^\xi_{n\lambda}&=\lambda\hbar\omega_B\sqrt{n+\gamma^2} \quad (n=1,2,3,\dots; \lambda=\pm1),\\
        \epsilon_0^\xi&=-\xi \Delta,
    \end{align}
\end{subequations}
where $\omega_B = v_F\sqrt{2eB/\hbar}$ is the effective cyclotron frequency and $\gamma = \Delta/(\hbar\omega_B)$ is the dimensionless gap. The index $\lambda = \pm 1$ distinguishes the electronlike and holelike branches. For the zeroth LL ($n=0$), valley--sublattice locking makes the energy $\epsilon_0^\xi$ depend explicitly on $\xi$ and thereby produces a valley-dependent response.

It is sufficient to evaluate the wave functions and observables in the $K$ valley ($\xi = +1$). The results for the $K'$ valley ($\xi = -1$) follow from $\Delta \to -\Delta$, as ensured by the unitary transformation
\begin{equation}
    H^{\xi}(\Delta)=\sigma_y H^{-\xi}(-\Delta)\sigma_y.
\end{equation}

With the bosonic ladder operator $\hat{a} = (\hat{\pi}_x - i\hat{\pi}_y)/\sqrt{2\hbar eB}$, the $K$-valley eigenstates $\ket{n\lambda}$ for $n > 0$ are expressed in terms of the harmonic-oscillator Fock states $\ket{n}$ as
\begin{equation}
    \label{n_lambda}
    \ket{n\lambda}=\begin{pmatrix} \sin\frac{\alpha_n^\lambda}{2}\ket{n-1} \\ \cos\frac{\alpha_n^\lambda}{2}\ket{n} \end{pmatrix},
\end{equation}
where the coefficients satisfy
\begin{equation}
    \label{sin_cos}
    \sin\alpha_n^\lambda=\lambda\sqrt{\frac{n}{n+\gamma^2}}, \quad \cos\alpha_n^\lambda=\frac{-\lambda\gamma}{\sqrt{n+\gamma^2}}.
\end{equation}
The zeroth LL ($n=0$) is fully polarized on one sublattice and has the form $\ket{0}=(0,\ket{0})^T$.

The viscosity calculation uses the microscopic stress tensor $\hat{T}_{ij}=\frac{v_F}{2}(\hat{\pi}_i\sigma_j+\hat{\pi}_j\sigma_i)$. In the ladder-operator basis, the relevant components are
\begin{equation}
    \label{T-matrix}
    \begin{aligned}
    \hat{T}_{xx}&=\frac{\hbar\omega_B}{2}
    \begin{pmatrix} 0 & \hat{a}+\hat{a}^\dagger \\ \hat{a}+\hat{a}^\dagger & 0 \end{pmatrix},\\
    \hat{T}_{xy}&=\frac{\hbar\omega_B}{2}
    \begin{pmatrix} 0 & -i\hat{a}^\dagger \\ i\hat{a} & 0 \end{pmatrix}.
    \end{aligned}
\end{equation}

The expressions below are given per Landau-level orbital and per spin. To obtain the 2D viscosity density, they must be multiplied by the orbital degeneracy $1/(2\pi l_B^2)$ and, when appropriate, by the spin degeneracy.

\subsection{Equivalence of the Two Formalisms}
\label{Equivalence of the Two Formalisms}
We first evaluate the Hall viscosity from the conventional stress--stress correlation function. Substituting the stress-tensor matrix elements in the ladder-operator basis and taking the dc limit, $\omega\to0$, gives
\begin{equation}
\begin{aligned}
\eta_H &= \frac{\hbar^3\omega_B^2}{2}\sum^{n>0}_{\lambda,\lambda^\prime=\pm1}
\cos^2\frac{\alpha_n^{\lambda}}{2}
\sin^2\frac{\alpha_{n+2}^{\lambda^\prime}}{2}
(n+1)F_{n\lambda,n+2\lambda^\prime} \\
&\quad + \sum_{\lambda=\pm1}\frac{\hbar^3\omega_B^2}{2}
\sin^2\frac{\alpha_2^\lambda}{2}F_{0,2\lambda}.
\end{aligned}
\end{equation}

At zero temperature, the Fermi--Dirac distribution reduces to a step function. If the highest occupied positive-energy LL has index $N\geq2$, the Kubo expression can be written as
\begin{equation}
\begin{aligned}
\label{T=0}
\eta_H
&=\frac{\hbar}{8}\sum_{n,\lambda}
(n+1)\left(n+1+\cos\alpha_n^\lambda\right)
\left[f(\epsilon_n^\lambda)-f(\epsilon_{n+2}^\lambda)\right] \\
&=\frac{\hbar}{8}\left[
N^2+(N-1)^2
-\frac{N\gamma}{\sqrt{N+\gamma^2}}
-\frac{(N-1)\gamma}{\sqrt{N-1+\gamma^2}}
\right].
\end{aligned}
\end{equation}
In this occupation-difference representation, transitions entirely within the filled negative-energy branch cancel because $f(\epsilon_n^-)-f(\epsilon_{n+2}^-)=0$. Equation~\eqref{T=0} is therefore finite without an explicit ultraviolet regularization.

Alternatively, one may first sum over the intra- and interband indices, $\lambda^\prime=\pm\lambda$, and formally rearrange the result into an occupied-state representation. Including the contribution from the zeroth LL gives
\begin{equation}
\label{eq:eta_H_exact}
\eta_H(\omega\to0)
=\frac{\hbar}{4}f(\epsilon_0)
+\frac{\hbar}{4}\sum_{n>0,\lambda}
\left(2n-\frac{\lambda\gamma}{\sqrt{n+\gamma^2}}\right)
f(\epsilon_{n\lambda}).
\end{equation}
Unlike Eq.~\eqref{T=0}, Eq.~\eqref{eq:eta_H_exact} is written as a sum over all occupied states. For the negative-energy branch, $f(\epsilon_{n,-})\to1$ as $n\to\infty$, and the Dirac-sea contribution is ultraviolet divergent. The passage from the occupation-difference form in Eq.~\eqref{T=0} to the occupied-state form in Eq.~\eqref{eq:eta_H_exact} is therefore only formal unless a prescription for the high-energy states is specified.

This subtlety is particularly important for an unbounded Dirac spectrum. With a finite number of LLs, the rearrangement is an exact telescoping identity. In the continuum Dirac model, however, sending the ultraviolet cutoff to infinity generates boundary terms from the filled negative-energy sea. The result then depends on how those terms are treated, for example by vacuum subtraction, a finite LL cutoff, or Pauli--Villars regularization. Consequently, the zero-temperature limit of Eq.~\eqref{eq:eta_H_exact} need not reproduce Eq.~\eqref{T=0} unless the same ultraviolet prescription and order of limits are used consistently.

We next evaluate the same response using the inhomogeneous-electric-field construction. To obtain the spatially dispersive coefficients, we introduce ladder operators for the guiding-center coordinates,
\begin{equation}
    \hat{b}\equiv\frac{1}{\sqrt{2}l_B}(\hat{X}+i\hat{Y}), \quad \hat{b}^\dagger\equiv\frac{1}{\sqrt{2}l_B}(\hat{X}-i\hat{Y}),
\end{equation}
where the coordinates decompose into guiding-center and relative cyclotron parts as $(\hat{x},\hat{y})\equiv(\hat{X},\hat{Y})+(\hat{R}_x,\hat{R}_y)$, with $(\hat{R}_x,\hat{R}_y)\equiv(\hat{\pi}_y,-\hat{\pi}_x)/(eB)$. To linear order in the complex wave vector $q\equiv q_x+iq_y$, the spatial phase factor is
\begin{equation}
    1+i\boldsymbol{q}\cdot\hat{\boldsymbol{r}}=1+\frac{il_B}{\sqrt{2}}\left[q^*\hat{b}+q\hat{b}^\dagger+i(q\hat{a}-q^*\hat{a}^\dagger)\right].
\end{equation}

The relevant matrix elements are expanded in powers of $q$ as
\begin{subequations}
    \begin{align}
        \langle m\lambda_m|\hat{T}_{xy}(\boldsymbol{q})|n\lambda_n\rangle&=[\hat{T}_{xy}]_{mn}^{(0)}+[\hat{T}_{xy}]_{mn}^{(1)}+\cdots, \\
        \langle n\lambda_n|\hat{v}_y(-\boldsymbol{q})|m\lambda_m\rangle&=[\hat{v}_y]_{nm}^{(0)}+[\hat{v}_y]_{nm}^{(1)}+\cdots,
    \end{align}
\end{subequations}
and the leading cross terms are
\begin{widetext}
\begin{subequations}
    \begin{align}
        [\hat{v}_y]_{nm}^{(0)}[\hat{T}_{xy}]_{mn}^{(1)}&=\frac{\hbar\omega_Bl_B}{2\sqrt{2}}\left[q\delta_{n+1,m}\sin^2\frac{\alpha_m^{\lambda_m}}{2}\cos^2\frac{\alpha_n^{\lambda_n}}{2}\left(n+\frac{1}{2}\right)-q^*\delta_{n,m+1}\sin^2\frac{\alpha_n^{\lambda_n}}{2}\cos^2\frac{\alpha_m^{\lambda_m}}{2}\left(m+\frac{1}{2}\right)\right], \\
        [\hat{v}_y]_{nm}^{(1)}[\hat{T}_{xy}]_{mn}^{(0)}&=\frac{\hbar\omega_Bl_B}{2\sqrt{2}}\left[q^*\delta_{n,m+2}\sin^2\frac{\alpha_n^{\lambda_n}}{2}\cos^2\frac{\alpha_m^{\lambda_m}}{2}(m+1)-q\delta_{n+2,m}\sin^2\frac{\alpha_m^{\lambda_m}}{2}\cos^2\frac{\alpha_n^{\lambda_n}}{2}(n+1)\right].
    \end{align}
\end{subequations}
\end{widetext}

When the Fermi level lies in a bulk gap between well-separated LLs, $\sigma_{xx}=0$. With the sign conventions used here, quantization of the Hall conductivity~\cite{p.m.krstajicIntegerQuantumHall2012,paoloceaQuantumHallEffect2012} reduces $\kappa_2=e\bar{n}\rho_{xy}$ to $\kappa_2=-B$. Substitution of the expanded matrix elements into the stress--velocity correlator reproduces the finite occupation-difference result in Eq.~\eqref{T=0}. The same algebra can be rearranged into Eq.~\eqref{eq:eta_H_exact} only after the same ultraviolet prescription is imposed.

\subsection{Physical Implications of the Valley Hall Viscosity}
\label{Physical Implications of the Valley Hall Viscosity}
The expressions above give the contribution from the $K$ valley ($\xi=+1$). The transformation $\gamma\to-\gamma$ determines the $K'$-valley response ($\xi=-1$). Therefore, only terms even in $\gamma$ survive in the valley-summed Hall viscosity, whereas the odd terms define the valley contrast.

The result separates into two contributions:
\begin{itemize}
    \item \textbf{Valley-independent 2DEG-like contribution:} The term proportional to $2n$ is identical in both valleys. It has the form of the Hall viscosity of a conventional two-dimensional electron gas, in which each fully occupied $n$th LL contributes $n\hbar/2$ per orbital~\cite{mohammadsherafatiHallViscosityElectromagnetic2016,mohammadsherafatiHallViscosityNonlocal2019}.
    \item \textbf{Valley-contrasting contribution:} The second term depends on the dimensionless gap $\gamma$ and changes sign between the $K$ and $K'$ valleys. This antisymmetric term is the valley-dependent contribution associated with the massive-Dirac band geometry~\cite{mohammadsherafatiHallViscosityNonlocal2019,selchValleyHallViscosity2026}.
\end{itemize}

Particle--hole symmetry (PHS) further constrains the response. Because the continuum Dirac spectrum is unbounded from below, the occupied-state representation requires an explicit regularization of the Dirac sea. The gap-dependent asymptotic sums can, for example, be expressed using the Hurwitz zeta function $\zeta(1/2,\gamma^2)$~\cite{claudioo.dibMagnetizedElectronGas2001,linasvepstasEfficientAlgorithmAccelerating2008}; the valley-independent vacuum contribution must be treated using the same prescription.

\begin{itemize}
    \item \textbf{Gapless limit ($\Delta\to0$):} Monolayer graphene has exact PHS within each valley. With a PHS-preserving vacuum subtraction, the neutrality-point Hall viscosity vanishes when all negative-energy LLs are filled and the zeroth LL is half filled. The half-filled zeroth level contributes $\hbar/8$ per orbital in Eq.~\eqref{eq:eta_H_exact}~\cite{thomasi.tuegelHallViscosityMomentum2015,tarokimuraHallSpinHall2021}.
    \item \textbf{Gapped regime ($\Delta\neq0$):} The single-valley LL spectrum is not particle--hole symmetric because valley--sublattice locking shifts the zeroth level to $E_0^\xi=-\xi\Delta$; PHS is recovered only after both valleys are included. For $|\epsilon_F|<|\Delta|$, the two valleys therefore have distinct regularized contributions. When the same set of nonzero LLs is occupied in both valleys, their sum reduces to the gapless result, whereas their difference remains gap dependent.
\end{itemize}

\section{Valley Hall Viscosity without a Magnetic Field}
\label{Valley Hall Viscosity without magnetic field}
We now consider the valley-resolved response in the absence of an external magnetic field. Hall viscosity is the antisymmetric, $\mathcal{T}$-odd part of the viscosity tensor, so a nonzero macroscopic charge response requires broken time-reversal symmetry. The inhomogeneous-electric-field construction separates the result into the spatially dispersive coefficient $\chi$, which is $\mathcal{T}$ even, and a transverse Hall response, which supplies the required $\mathcal{T}$-odd factor. In a multivalley system with weak intervalley relaxation, the response can remain finite within each valley while canceling in the valley-summed channel, as in the valley Hall effect.

\subsection{Model and Linear Response}
\label{Model and Linear Response}
The low-energy effective Hamiltonian for gapped graphene at zero magnetic field is
\begin{equation}
    \label{modelB=0}
    h_0^\xi(\boldsymbol{k})=\hbar v_F(\xi k_x\sigma_x+k_y\sigma_y)+\Delta\sigma_z.
\end{equation}
For the $K$ valley ($\xi=+1$), momentum derivatives of the Hamiltonian define the microscopic stress tensor and velocity operator~\cite{hassanshapourianViscoelasticResponseTopological2015,pranavraoHallViscosityQuantum2020}:
\begin{equation}
    \label{stressnew}
    \hat{T}_{ij}=\frac{1}{2}\left(k_i\frac{\partial h_0}{\partial k_j}+k_j\frac{\partial h_0}{\partial k_i}\right), \quad \hat{v}_k=\frac{1}{\hbar}\frac{\partial h_0}{\partial k_k}.
\end{equation}

A vector potential $\boldsymbol{A}(\boldsymbol{r},t)$ represents the inhomogeneous electric field through $\boldsymbol{E}(\boldsymbol{r},t)=-\partial_t\boldsymbol{A}(\boldsymbol{r},t)$ and generates the perturbation $H^\prime=e\boldsymbol{A}(\boldsymbol{r},t)\cdot\hat{\boldsymbol{v}}$. Linear-response theory gives $T_{ij}(\boldsymbol{q},\omega)=G_{ijk}(\boldsymbol{q},\omega)A_k(\boldsymbol{q},\omega)$, with
\begin{equation}
    \label{G1}
    \begin{aligned}
        G_{ijk}(\boldsymbol{q},\omega) &= \frac{e}{2i\hbar}\sum_{m,n,\boldsymbol{k}} \frac{f[\varepsilon_m(\boldsymbol{k}_-)] - f[\varepsilon_n(\boldsymbol{k}_+)]}{\varepsilon_m(\boldsymbol{k}_-) - \varepsilon_n(\boldsymbol{k}_+) + i\hbar\omega^+} \\
        &\quad \times \left[k_i F_{jk}^{mn}(\boldsymbol{k},\boldsymbol{q}) + k_j F_{ik}^{mn}(\boldsymbol{k},\boldsymbol{q})\right],
    \end{aligned}
\end{equation}
where $\boldsymbol{k}_\pm=\boldsymbol{k}\pm\boldsymbol{q}/2$, $\varepsilon_n(\boldsymbol{k})$ is the band energy, and $F_{jk}^{mn}(\boldsymbol{k},\boldsymbol{q})$ is the generalized current--current matrix element
\begin{equation}
    \label{F}
    F_{jk}^{mn}(\boldsymbol{k},\boldsymbol{q}) = \langle u_m(\boldsymbol{k}_-) | \hat{v}_j | u_n(\boldsymbol{k}_+) \rangle \langle u_n(\boldsymbol{k}_+) | \hat{v}_k | u_m(\boldsymbol{k}_-) \rangle,
\end{equation}
where $\ket{u_n(\boldsymbol{k})}$ is the periodic part of a Bloch state.

The coefficient $\chi$ follows from the limit
\begin{equation}
    \label{munew}
    \chi=\lim_{\boldsymbol{q}\to0}\lim_{\omega\to0}\frac{1}{\omega}\frac{\partial}{\partial q_x}G_{xyy}(\boldsymbol{q},\omega).
\end{equation}
The order of limits corresponds to taking the quasistatic limit at finite $\boldsymbol{q}$ before sending $\boldsymbol{q}$ to zero. We therefore expand Eq.~\eqref{G1} through order $\mathcal{O}(q\omega)$.

\subsection{Systems with Particle--Hole Symmetry}
\label{Systems with Particle-Hole Symmetry}
We first place the chemical potential $\mu\leq -|\Delta|$ in the valence band. For exact particle--hole symmetry (PHS), $\varepsilon_V(\boldsymbol{k})=-\varepsilon_C(\boldsymbol{k})$, where V and C label the valence and conduction bands. The response then reduces to a Fermi-surface contribution:
\begin{equation}
    \label{4.12}
    \begin{aligned}
        &\frac{1}{\omega^+}G_{ijk}(\boldsymbol{q},\omega) \\&= -\frac{e}{2}\sum_{\boldsymbol{k}}\delta(\varepsilon_V(\boldsymbol{k})-\mu)(\hbar\boldsymbol{v}_V(\boldsymbol{k})\cdot \boldsymbol{q})\ k_i g_{jk}(\boldsymbol{k})+(i\leftrightarrow j)\\
        &= -\frac{e}{2}\int\frac{\mathrm d S}{|\boldsymbol{v}_V(\boldsymbol{k})|}(\hbar\boldsymbol{v}_V(\boldsymbol{k})\cdot \boldsymbol{q})\ k_i g_{jk}(\boldsymbol{k})+(i\leftrightarrow j),
    \end{aligned}
\end{equation}
where $S$ denotes the Fermi surface. In the units used in the following formulas, the valence-band group velocity is $\boldsymbol{v}_V(\boldsymbol{k})=-\boldsymbol{k}/\varepsilon(\boldsymbol{k})$, and the quantum metric is~\cite{shunjimatsuuraMomentumSpaceMetric2010}
\begin{equation}
    \label{4.13}
    g_{ij}(\boldsymbol{k})=\frac{1}{4\varepsilon^4(\boldsymbol{k})}\left[\delta_{ij}\varepsilon^2(\boldsymbol{k})-k_ik_j\right],
\end{equation}
with $\varepsilon(\boldsymbol{k})=\sqrt{k^2+\Delta^2}$. Equation~\eqref{4.12} depends only on the Fermi surface.

\begin{figure*}[t]
  \centering
  \includegraphics[width=0.95\textwidth]{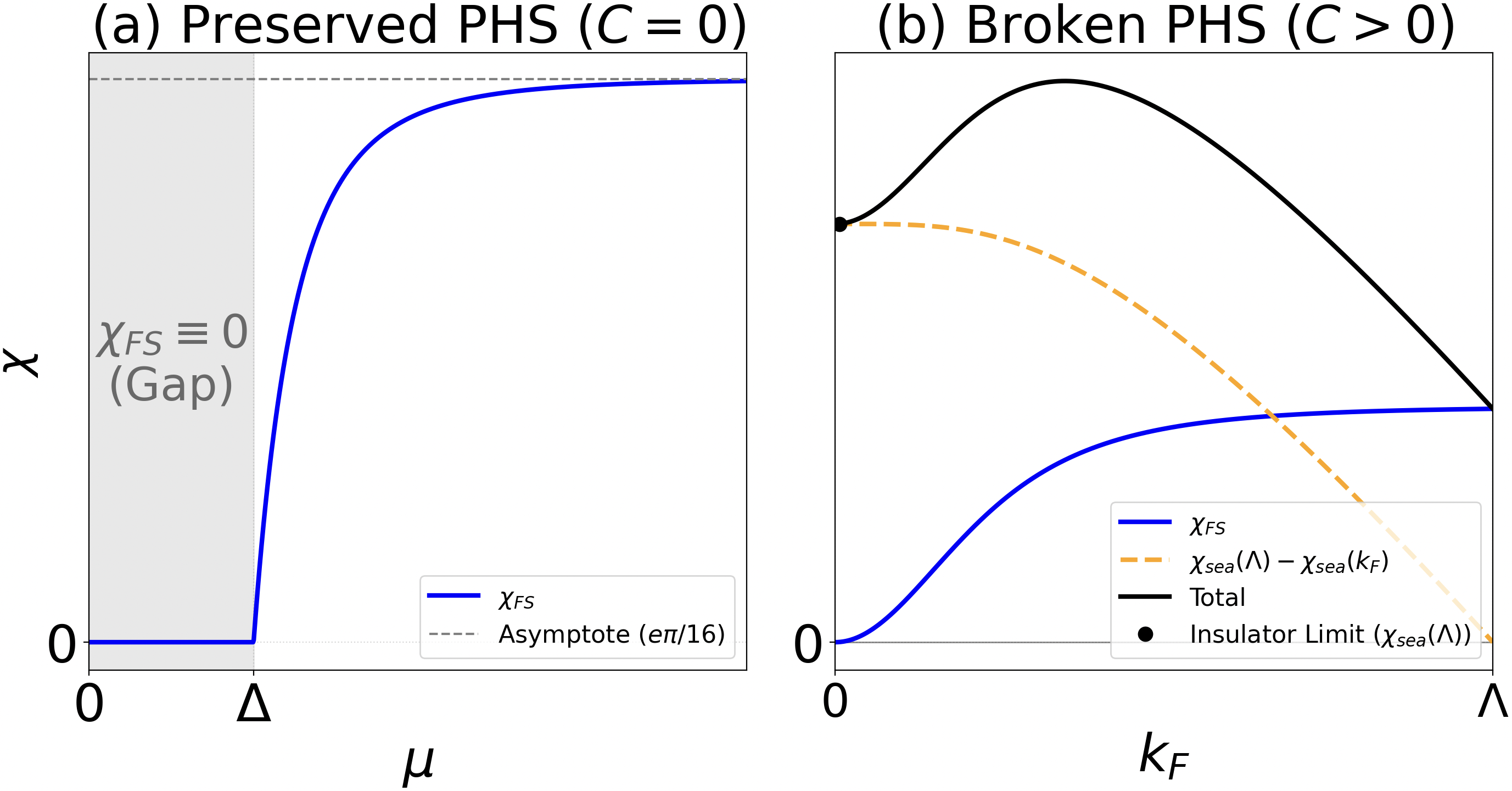}
  \caption{Dependence of $\chi$ on the chemical potential $\mu$ and Fermi momentum $k_F$. (a) Fermi-surface contribution $\chi_{\text{FS}}$ in the particle--hole-symmetric (PHS) case ($C=0$). The response vanishes when the chemical potential lies in the band gap ($\mu<|\Delta|$, gray region) and approaches $e\pi/16$ at high doping. (b) PHS-broken case ($C>0$). For the parameters shown, the occupied valence-band states within the cutoff contribute both a Fermi-surface term (blue line) and a Fermi-sea term (dashed line). The black curve is their sum. The black dot denotes the insulating limit ($k_F=0$), in which the Fermi-surface term vanishes and the total response equals $\chi_{\text{sea}}(\Lambda)$.}
  \label{fig:chi}
\end{figure*}

For this system, Eq.~\eqref{munew} yields
\begin{equation}
    \label{4.14}
    \chi_{\text{FS}}=\frac{e\pi}{16}\left(1-\frac{\Delta^4}{\mu^4}\right).
\end{equation}
When the chemical potential lies in the gap, the Fermi surface is absent and Eq.~\eqref{4.12} gives $\chi=0$ [Fig.~\ref{fig:chi}(a)].

The corresponding Hall conductivity is~\cite{dixiaoValleyContrastingPhysicsGraphene2007}
\begin{equation}
    \label{4.15}
    \sigma_H=\frac{e^2}{2h}\frac{\Delta}{\mu}.
\end{equation}
Combining this conductivity with the spatially dispersive response gives the valley-resolved Hall coefficient
\begin{equation}
    \label{4.16}
    \eta_H=e\bar{n}\frac{\sigma_H}{\sigma_{xx}^2+\sigma_H^2}\chi,
\end{equation}
where $\bar{n}$ is the carrier density assigned to the independently conserved valley channel. Equation~\eqref{4.16} assumes that intervalley relaxation is negligible and that the resistivity is defined for that channel; it is not obtained by inverting the total charge conductivity, whose Hall component vanishes at zero magnetic field.

Valley exchange, $\Delta\to-\Delta$, maps the response to the $K'$ valley and gives $\eta_H^{K'}=-\eta_H^K$. Global time-reversal symmetry therefore enforces $\eta_H^{\mathrm{total}}=\eta_H^K+\eta_H^{K'}=0$. When the two valley channels are well defined, their difference, $\eta_H^v\equiv\eta_H^K-\eta_H^{K'}$, defines a valley Hall viscosity analogous to the valley Hall conductivity generated by valley-contrasting Berry curvature. In the present construction, the quantum-metric coefficient $\chi$ is valley even, whereas the transverse Hall factor is valley odd.

\subsection{Role of PHS Breaking: Fermi Sea Contributions}
\label{Role of PHS Breaking: Fermi Sea Contributions}
Exact PHS forces $\chi$ to vanish in the insulating state, where no Fermi surface is present. We model PHS breaking by adding the leading isotropic scalar term $Ck^2\sigma_0$ to Eq.~\eqref{modelB=0}:
\begin{equation}
    \label{modelB=0_PHS}
    h_C^\xi(\boldsymbol{k})=h_0^\xi(\boldsymbol{k})+Ck^2\sigma_0.
\end{equation}
In a lattice description of graphene, such a term arises from next-nearest-neighbor hopping.

\begin{figure}[t]
  \centering
  \includegraphics[width=\columnwidth]{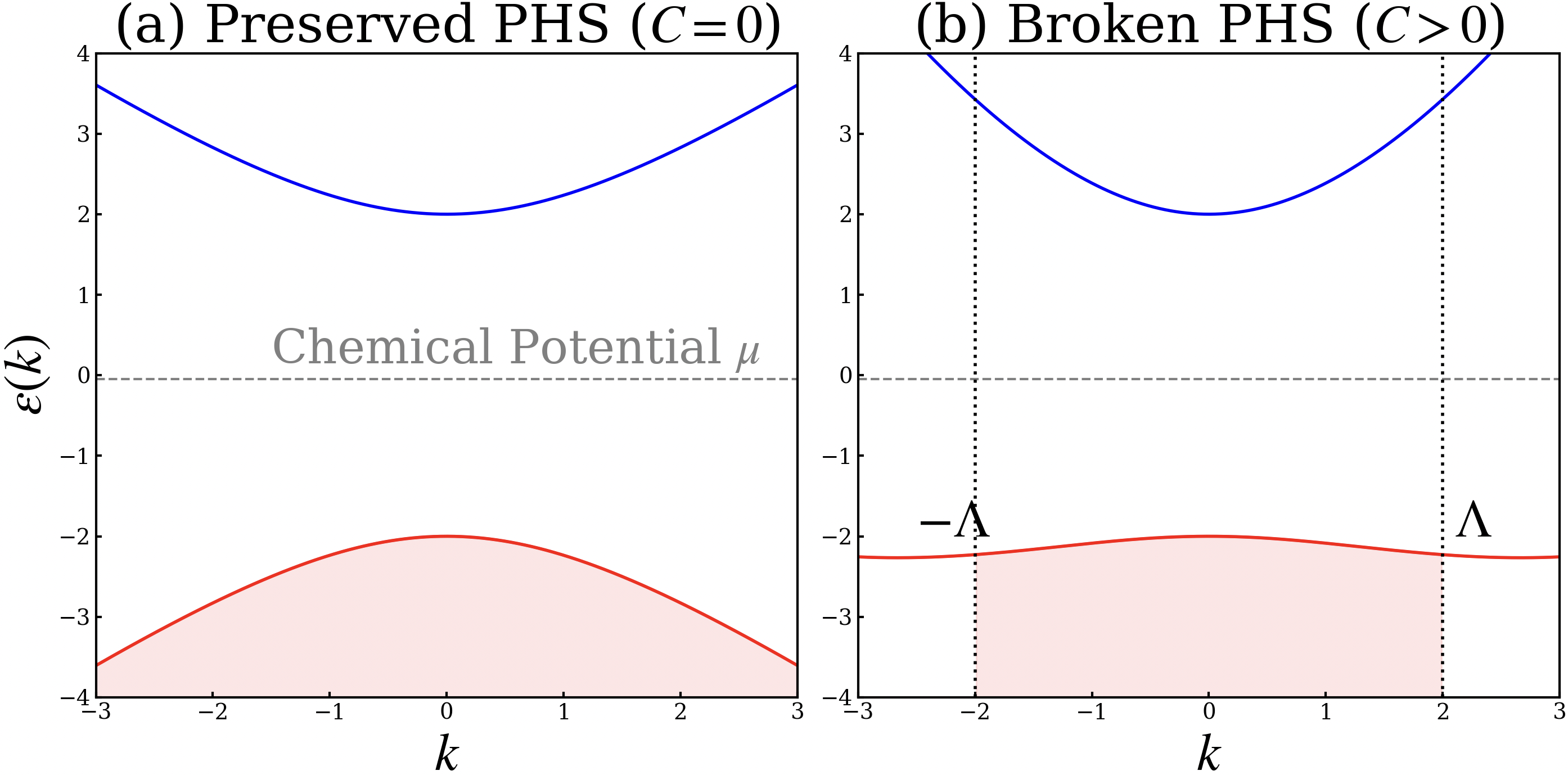}
  \caption{Band dispersion and Fermi-sea regularization when the chemical potential $\mu$ lies in the gap. (a) With exact particle--hole symmetry ($C=0$), the conduction ($\varepsilon_C$) and valence ($\varepsilon_V$) bands are symmetric. (b) When the $Ck^2$ term breaks particle--hole symmetry ($C\Delta/v_F^2=0.3$), the continuum valence band eventually bends upward at large momentum and crosses the chemical potential. The ultraviolet cutoff $\Lambda$ restricts the model to its low-energy domain (shaded region) and preserves the insulating filling used in the calculation.}
  \label{fig:PHS}
\end{figure}

After PHS is broken, the average band velocity is $\bar{\boldsymbol{v}}(\boldsymbol{k})=\hbar^{-1}\nabla_{\boldsymbol{k}}(Ck^2)=2C\boldsymbol{k}/\hbar$ rather than zero. This generates a geometric Fermi-sea contribution in addition to any Fermi-surface term. In the insulating state, the linear-response expansion contains
\begin{equation}
    \label{eq:G_PHS_breaking}
    \begin{aligned}
        &\frac{1}{\omega^+}G_{ijk}(\boldsymbol{q},\omega)\\ &= e\sum_{\boldsymbol{k}}\frac{k_i\Theta(\Lambda-k)}{\varepsilon_C(\boldsymbol{k})-\varepsilon_V(\boldsymbol{k})} \Big\{2\hbar\bar{\boldsymbol{v}}(\boldsymbol{k})\cdot \boldsymbol{q}\ g_{jk}(\boldsymbol{k}) \\
        &\quad - \left[\hbar\bar{v}_j(\boldsymbol{k})\ g_{kl}(\boldsymbol{k})+\hbar\bar{v}_k(\boldsymbol{k}) \ g_{jl}(\boldsymbol{k})\right]q_l\Big\} + (i\leftrightarrow j).
    \end{aligned}
\end{equation}
The unbounded $Ck^2$ term makes the valence band $\varepsilon_V(\boldsymbol{k})$ bend upward at large $|\boldsymbol{k}|$. Without regularization, the band eventually crosses the chemical potential and makes the continuum model spuriously metallic (Fig.~\ref{fig:PHS}). We therefore introduce an ultraviolet (UV) momentum cutoff $\Lambda$ that represents the finite validity range of the Dirac expansion rather than a literal circular Brillouin-zone boundary. For a chemical potential $\mu$ in the gap, the regularized model is insulating provided
\begin{equation}
    \max_{k\leq\Lambda}\varepsilon_V(k)<\mu<\min_{k\leq\Lambda}\varepsilon_C(k).
\end{equation}

In this insulating regime, all valence-band states within the cutoff are occupied. Substituting the Dirac quantum metric and the average velocity $\bar{\boldsymbol{v}}(\boldsymbol{k})$ into Eq.~\eqref{eq:G_PHS_breaking} reduces $\chi$ to an integral over the regularized Fermi sea:
\begin{equation}
    \label{eq:mu_integral}
    \begin{aligned}
        \chi_{\text{sea}} &= \frac{e C}{4}\int_{k<\Lambda}\mathrm{d^2}\boldsymbol{k}\frac{k_x^2}{(k^2+\Delta^2)^{3/2}} \\
        &= \frac{e C \pi}{4} \left( \sqrt{\Lambda^2+\Delta^2} + \frac{\Delta^2}{\sqrt{\Lambda^2+\Delta^2}} - 2|\Delta| \right).
    \end{aligned}
\end{equation}
The cutoff $\Lambda$ removes the linear UV divergence of the unbounded continuum integral. The resulting coefficient is a nonuniversal band-structure contribution that must ultimately be fixed by a lattice regularization.

As shown in Fig.~\ref{fig:chi}(b), the total $\chi$ of a PHS-broken system contains both Fermi-sea and Fermi-surface contributions. In the insulating limit, only the Fermi-sea term remains. Upon doping, the two contributions coexist, with their relative magnitude depending on $C$, $\Delta$, the chemical potential, and the cutoff.

\section{Conclusion}
\label{Conclusion}
We have calculated the valley-dependent Hall response of gapped two-dimensional Dirac fermions using a viscoelastic formulation based on static, spatially inhomogeneous electric fields. In a perpendicular magnetic field, an explicit Landau-level calculation reproduces the conventional strain-based Kubo result and reveals a valley-asymmetric contribution. At zero magnetic field, global time-reversal symmetry ($\mathcal{T}$) forces the valley-summed Hall viscosity to vanish, whereas the two valley-resolved responses are equal and opposite. Within the present construction, the quantum-metric coefficient is valley even and the transverse Hall factor is valley odd. In a fully gapped insulator, exact particle--hole symmetry eliminates the zero-field response; a quadratic particle--hole-asymmetric term generates a cutoff-dependent Fermi-sea contribution. The latter is nonuniversal within the continuum model and must ultimately be fixed by a lattice regularization. These results clarify how valley-dependent viscoelasticity can be encoded in electromagnetic response. Establishing a direct experimental protocol will require a valley-resolved preparation or detection scheme and an explicit treatment of intervalley relaxation.

We thank Takahiro Uto for insightful discussions on detection schemes for our system. This work was supported by the National Natural Science Foundation of China (NSFC) under Grant No. 12374126, 
by the Priority Program of Chinese Academy of Sciences under Grant No. XDB28000000, 
and by JSPS KAKENHI for Grants (Nos. 23H01839, 24H00322, 24H00853, and 25K07224) from MEXT, Japan.

\onecolumngrid 
\appendix 
\section{Perturbation Expansion}
\label{Perturbation Expansion}
At zero temperature, with the chemical potential $\mu$ in the valence band, the interband contribution to Eq.~\eqref{G1} is
\begin{equation}
    \label{G2}
    \begin{aligned}
        G_{ijk}(\boldsymbol{q},\omega) &= \frac{e}{2i\hbar}\sum_{\boldsymbol{k}} \frac{k_i f[\varepsilon_V(\boldsymbol{k}_-)] F_{jk}^{VC}(\boldsymbol{k},\boldsymbol{q})}{\varepsilon_V(\boldsymbol{k}_-) - \varepsilon_C(\boldsymbol{k}_+) + i\hbar\omega^+} \\
        &\quad - \frac{k_i f[\varepsilon_V(\boldsymbol{k}_+)] F_{jk}^{CV}(\boldsymbol{k},\boldsymbol{q})}{\varepsilon_C(\boldsymbol{k}_-) - \varepsilon_V(\boldsymbol{k}_+) + i\hbar\omega^+} + (i \leftrightarrow j),
    \end{aligned}
\end{equation}
where V and C denote the valence and conduction bands of Eq.~\eqref{modelB=0}, respectively. We evaluate Eq.~\eqref{G2} by expanding in the external frequency $\omega$ and wave vector $\boldsymbol{q}$ through order $\mathcal{O}(\omega q)$. The energy denominator becomes
\begin{align}
    \label{D}
    \frac{1}{\varepsilon_V(\boldsymbol{k}_-)-\varepsilon_C(\boldsymbol{k}_+)+i\hbar\omega^+}=-\frac{1}{\Delta\varepsilon(\boldsymbol{k})}+\frac{-i\hbar\omega^++\hbar\bar{\boldsymbol{v}}(\boldsymbol{k})\cdot\boldsymbol{q}}{\Delta\varepsilon(\boldsymbol{k})^2}+\frac{2i\hbar^2\omega^+\bar{\boldsymbol{v}}(\boldsymbol{k})\cdot\boldsymbol{q}}{\Delta\varepsilon(\boldsymbol{k})^3},
\end{align}
where $\boldsymbol{v}_n(\boldsymbol{k})=\hbar^{-1}\nabla_{\boldsymbol{k}}\varepsilon_n(\boldsymbol{k})$ denotes the band-velocity convention used here, $\Delta\varepsilon(\boldsymbol{k})=\varepsilon_C(\boldsymbol{k})-\varepsilon_V(\boldsymbol{k})$ is the direct gap, and $\bar{\boldsymbol{v}}(\boldsymbol{k})=\frac{1}{2}[\boldsymbol{v}_V(\boldsymbol{k})+\boldsymbol{v}_C(\boldsymbol{k})]$ is the band-averaged velocity. For a particle--hole-symmetric massive Dirac dispersion, $\varepsilon_V(\boldsymbol{k})=-\varepsilon_C(\boldsymbol{k})$, and hence $\bar{\boldsymbol{v}}(\boldsymbol{k})\equiv0$.

At zero temperature, the Fermi--Dirac distribution in Eq.~\eqref{G2} expands about $\boldsymbol{k}$ as
\begin{align}
    \label{f}
    f[\varepsilon_V(\boldsymbol{k} \mp \boldsymbol{q}/2)] \approx \Theta(k_F - k) \pm \frac{\hbar \boldsymbol{v}_V(\boldsymbol{k})\cdot\boldsymbol{q}}{2}\delta(\varepsilon_V(\boldsymbol{k}) - \mu),
\end{align}
where $\Theta(x)$ is the Heaviside step function.

Expansion of the interband matrix elements gives
\begin{align}
\label{F_expand}
F_{jk}^{VC}(\boldsymbol{k},\boldsymbol{q})=\Delta\varepsilon(\boldsymbol{k})^2Q^V_{jk}(\boldsymbol{k})+\Delta\varepsilon(\boldsymbol{k})\cdot\big[\hbar\bar{v}_j(\boldsymbol{k})Q^V_{lk}+\hbar\bar{v}_k(\boldsymbol{k})Q^V_{jl}\big]q_l,
\end{align}
where $Q^V_{ij}(\boldsymbol{k})=\langle \partial_{k_i}u_V|(1-|u_V\rangle\langle u_V|)|\partial_{k_j}u_V\rangle$ is the quantum geometric tensor (QGT) of the valence band. Its symmetric real part is the quantum metric $g_{ij}$, whereas its antisymmetric imaginary part is the Berry curvature $\Omega_{ij}$; hence, $Q^V_{ij}=g_{ij}-\frac{i}{2}\Omega_{ij}$. For this two-band model, the conduction-band tensor is the complex conjugate of the valence-band tensor, $Q^C_{ij}(\boldsymbol{k})=[Q^V_{ij}(\boldsymbol{k})]^*$.\cite{tomokiozawaRelationsTopologyQuantum2021,mitscherlingGaugeinvariantProjectorCalculus2025}

\twocolumngrid

\bibliography{ref}

\end{document}